\documentclass[twocolumn,showpacs,preprintnumbers,amsmath,amssymb,aps,prl,groupedaddress]{revtex4-1}

\pdfoutput=1

\usepackage{amsmath,amssymb}
\usepackage{bm}
\usepackage{graphicx,color,natbib}
\usepackage{placeins}
\usepackage[dvipsnames]{xcolor}

\newcommand{\beq}{\begin{equation}}
\newcommand{\eeq}{\end{equation}}
\newcommand{\beqa}{\begin{eqnarray}}
\newcommand{\eeqa}{\end{eqnarray}}

\definecolor{strawberry}{rgb}{1.0,0.0,0.5}

\newcommand{\pecl}{\operatorname{\mathit{P\kern-.08em e}}}

\begin{document}
\title{
Active Dipolar Colloids in Three Dimensions: Strings, Sheets, Labyrinthine Textures and Crystals}

\author{Nariaki Saka\"i}
\email{nariaki.sakai@gmail.com}
\author{C. Patrick Royall}
\affiliation{H.H. Wills Physics Laboratory, Tyndall Avenue, Bristol, BS8 1TL, United Kingdom, School of Chemistry, University of Bristol, Cantock's Close, Bristol, BS8 1TS, United Kingdom, and Centre for Nanoscience and Quantum Information, Tyndall Avenue, Bristol, BS8 1FD, United Kingdom}
\email{paddy.royall@bristol.ac.uk}

\date{\today}

\begin{abstract}
Active matter exhibits striking behaviour reminiscent of living matter and molecular fluids, and has promising applications in drug delivery or mixing at the micron scale. Active colloidal systems provide important models with simple and controllable interactions amenable to theory and computer simulation. Experimental work is dominated by (quasi) two--dimensional (2d) systems at relatively low concentration, and rather less is known of the 3d case at concentrations pertinent to motility induced phase separation or to mimic morphogenesis. Here we investigate a 3d experimental system of active colloids in a dense suspension up to volume fractions of 0.5. The particles in our system are self-propelled in the lateral plane  under an AC electric field. The field in addition induces an electric dipole, and the competition between activity and both steric and dipolar interactions gives rise to phase behaviour ranging from an active gas to a dynamic labyrinthine phase as well as dense tetragonal and hexagonal crystals. Intermediate volume fractions are characterised by two--dimensional sheets with large fluctuations reminiscent of active membranes. These active sheets break symmetry in a direction perpendicular to the applied field. Moreover, the relationship between electric field and the particle dynamics depends in a complex and unexpected manner upon the position in the state diagram.
\end{abstract}


\maketitle

\textit{Introduction --- }
From bacteria \cite{Polin2009, Zhang2010,Jepson2013,Lopez2015}, biological tissues \cite{Czajkowski2018, Petrolli2019}, actin filaments \cite{Sanchez2012, Guillamat2017} and self-propelled colloids \cite{Palacci2010,Buttinoni2013,Palacci2013,Bricard2013} to shaken granular particles \cite{Narayan2007,Deseigne2010}, midges \cite{Attanasi2014} and birds \cite{Ballerini2008}, active matter is characterised by a broad range of length scales and systems. They are all characterised by an injection of energy which is then converted to mechanical motion. In other driven systems like sheared granular suspensions \cite{Sakai2019} or turbulent flows \cite{Michel2017}, energy is injected from the boundaries at a macroscopic level, flows towards the bulk and is dissipated at the microscopic level. On the contrary, energy in active systems is injected in the bulk at the particle level \cite{Ramaswamy2010,Marchetti2013,Elgeti2015,Bechinger2016,Zottl2016,Needleman2017,Zhang2017}. The resulting mechanical motion is often translational but may also be rotational \cite{Grzybowski2000, Fily2012, Nguyen2014, Yeo2015, Spellings2015, VanZuiden2016, Aubret2018}, and the direction of this motion is only determined by the particle orientation and not imposed by an external field. These properties make active matter rich and complex, and synthetic self-propelled particles are promising not only to mimic and understand living matter, but also to develop new material having novel response to external stimuli and potential application like mixing at the microscale \cite{Thampi2016}, drug delivery \cite{Baylis2015}, micromotors \cite{DiLeonardo2010, Krishnamurthy2016} or shape shifting \cite{Stenhammar2016, Frangipane2018, Arlt2018}.

\textit{Outstanding questions --- }
So far, much experimental effort has been focused on (quasi) two-dimensional (2d) systems, leading to the discovery of a complex phase and dynamical behaviour \textit{e.g.} a hexatic phase \cite{Cugliandolo2017}, clustering \cite{Peruani2006, Hagan2013, Cates2015}, rheotaxis \cite{Palacci2015}, self-assembly\cite{SchwarzLinek2012, Aubret2018, Mallory2018, Lowen2018} or mesoscopic turbulence \cite{Sanchez2012, Guillamat2017, Wensink2012}, and there remain many outstanding questions in the field. Nevertheless, the third dimension is essential to understanding many biological phenomena like morphogenesis of cell tissues \cite{Aman2010}, not to mention the profound difference between two- and three-dimensions in phase transitions \cite{Mognetti2013, Wysocki2014, Stenhammar2014, Wueaal2017, Shendruk2018, Metselaar2019}.

To this end, we consider a 3d active colloidal system in which the motility is in the lateral $xy$ plane. In addition to anisotropic activity, the particles feature a dipolar interaction due to the external electric field. In the case of passive colloids with dipolar interactions aligned with the field, the phase behaviour is well known \cite{Yethiraj2003, Hynninen2005}: at low volume fraction, particles assemble into strings aligned along the field direction, higher density leads to a phase behaviour of crystals, including body-centered tetragonal, BCC and FCC. By analogy then, we might expect the phase behaviour we encounter to be similar to a passive system, albeit with the possibility of some active phenomena \textit{e.g.} motility-induced phase separation (MIPS) that computer simulations and theory showed to also occur in three dimensions \cite{Cates2013, Stenhammar2014, Wysocki2014,Winkler2015}. Compared to such expectations, our system presents three unexpected findings. Firstly, we find a symmetry-breaking in the $xy$ plane which is not found in the passive system: the strings formed at low volume fraction self-assemble into two-dimensional \emph{sheets}, which fluctuate continuously due to the activity. Secondly, upon further increase of the volume fraction, rather than MIPS, the sheets percolate to form \emph{a dynamic labyrinthine phase}. Here the symmetry is broken along the field direction (in which the structure is replicated) while in the plane perpendicular to the field it is an active network which branches continuously and regenerates leading to an evolving structure. Thirdly, unlike many systems, we find that the dynamics of our system \emph{accelerate} upon crystallisation.

\begin{figure*}
\centering
\includegraphics[width=0.75\textwidth]{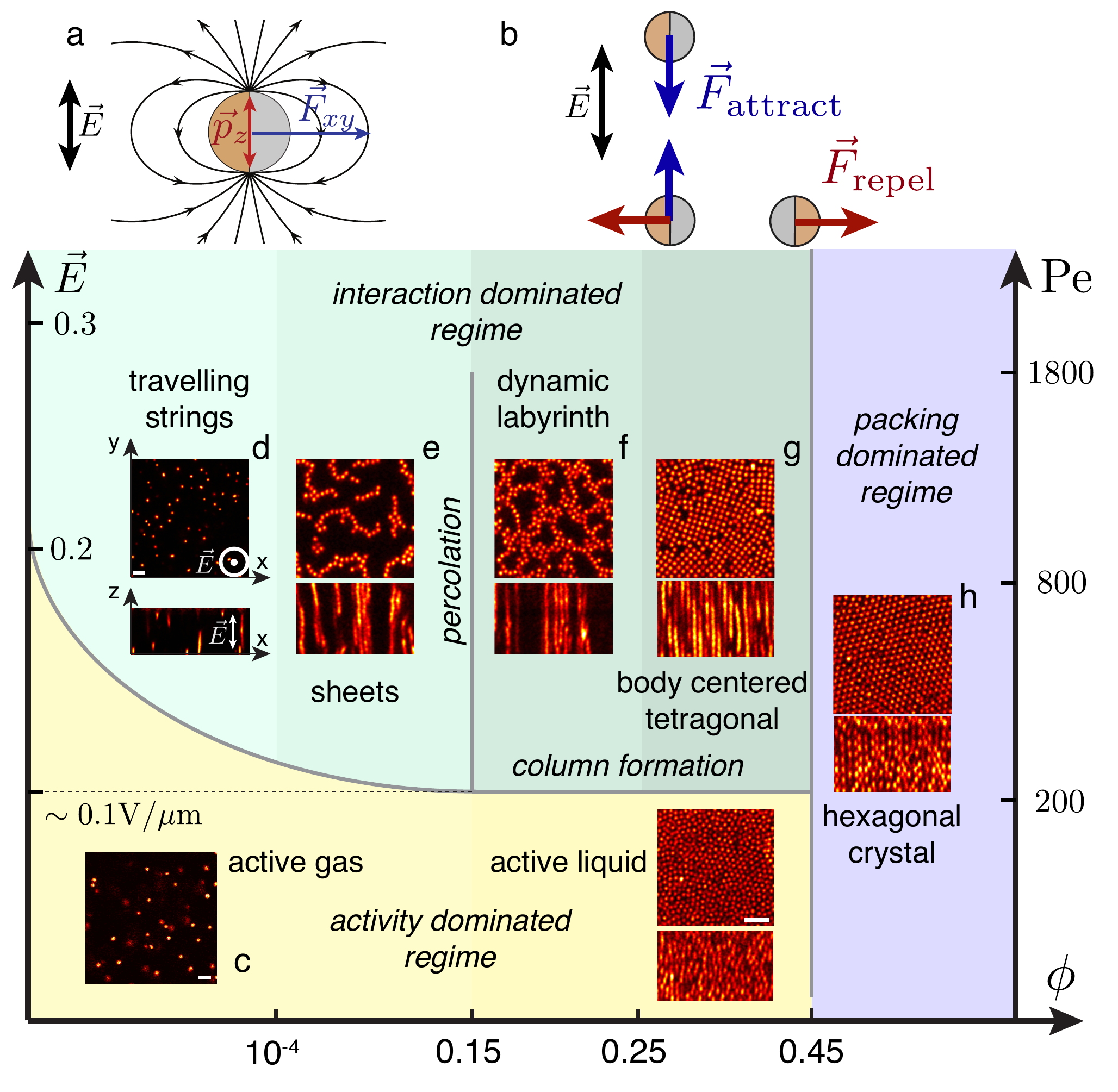}
\caption{\textit{State Diagram --- }
(a) The external AC electric field $\vec{E}$ induces an active force $\vec{F}_{xy}$ and electric dipole $\vec{p}_z$. The resulting field leads to dipolar interactions between the particles. (b) This interaction is anisotropic: two particles aligned with the electric field attract each other, whereas when they are perpendicular, the interaction is repulsive. There is a competition between the dipolar interaction and activity, which leads to a complex phase behaviour observed in our experiments. 
(c) At low field, the dipolar interaction is negligible and the suspension is in an active fluid state. Increasing the field leads to hierarchical auto-assembly into 1d strings 
(d) 2d sheets 
(e) and 3d labyrinthine phase (f) or tetragonal crystal (g), depending on the volume fraction, and these objects possess astonishing dynamical behaviour. At the highest volume fraction, packing dominates; the suspension crystallises and adopts hexagonal structure whatever the electric field (g). Images in the state diagram are two dimensional slices of 3d images obtained from our confocal microscope, and the P\'eclet number is computed using  $\mathrm{Pe} = \frac{v}{\sqrt{D_rD_t}}$, where $D_r$ and $D_t$ are respectively the rotational and translational diffusion coefficients of a single particle in the dilute limit, and where $v$ is the theoretical velocity of a single particle \cite{Gangwal2008} (see Supplementary Information, SI). Scale bars correspond to 5 microns.}
\label{figPhaseDiagram}
\end{figure*}

\textit{Experimental system --- }
The active colloids are self-propelled by induced-charge electrophoresis \cite{Gangwal2008,Nishiguchi2015,Yan2016,VanDerLinden2019}. We used two experimental systems depending on the range of volume fraction we studied. For higher and intermediate volume fractions, the particles are $1.5\mu m$ fluorescent silica particles (Kisker Biotech), one hemisphere of which is covered with $3\mathrm{nm}$ of Chromium using a thermal evaporator, then by an another$~15\mathrm{nm}$ shell of silica following \cite{Graf2003}. The thickness of the metal is chosen such that the layer remains transparent. The particles are suspended in a mixture of milliQ water and DMSO at volume ratio 7:10 for refractive index matching. For low volume fraction samples ($\phi$ less than $0.05$), $1\mu m$ fluorescent polystyrene particles (Invitrogen) are covered with 5nm of Aluminium, then 5nm of silica. The solvent is a mixture of water and glycerol at volume ratio 1:1, which allows us to density match the particles with the solvent and increase the viscosity.

The suspension is loaded into a cell formed by two ITO cover slips (SPI Supply) spaced by a distance $h\approx40\mathrm{\mu m}$, and an AC electric field is applied (Black star Jupiter 2010) at a frequency of $5\mathrm{kHz}$ (unless otherwise indicated) at different voltage. The system is imaged with a Leica TCS SP8 confocal microscope, and particles are tracked in space using ref. \cite{Leocmach2013}, and trajectories are reconstructed following \cite{linkingCode, Vercauteren2009}. For samples with polystyrene particles, the lack of index matching does not allow us to obtain particle trajectories, but we measured average velocity by manually recording individual displacements of ten particles between two successive frames.

The particles are self-propelled in the presence of an AC electric field, where the net active force is parallel to the electrodes [Fig.~\ref{figPhaseDiagram}(a)] \cite{Gangwal2008,Nishiguchi2015,Yan2016,VanDerLinden2019, Yethiraj2003}. Consequently, isolated particles move similarly to active Brownian particles in a plane orthogonal to the field ($xy$) and diffuse in the third direction ($z$). Moreover, due to the imbalance of the dielectric constant between the solvent and the particles, the local electric field is distorted, and the resulting interactions between the particles, can be modelled by two effective electric dipoles, one in each hemisphere of different magnitude, reflecting the different materials from which the two hemispheres of the Janus particles are comprised \cite{Stoy1994}. The dipolar interactions in this system have been carefully characterised and have a complex dependence on the frequency of the applied electric field \cite{Yan2016}. For our parameters (5 kHz and NaCl at a concentration of $10^{-4}\mathrm{mol/L}$), both dipoles have similar strengths, and so for simplicity we treat our particles as a single effective dipole located at the centre. Thus the interaction between two particles depends on their orientation: they attract when aligned with the field, and repel if orthogonal [Fig.~\ref{figPhaseDiagram}(b)]. This is reminiscent of some active dipole experiment and simulations \cite{Vincenti2019, Kaiser2015, Guzman-Lastra2016}, but in contrast to these works, activity and dipoles are in orthogonal directions.

The energy scale associated with two plain silica particles under a typical $0.1\mathrm{V/\mu m}$ for our experimental conditions is $U_d = 2.5 \times 10^{3}k_BT$ \cite{Stoy1994}, three orders of magnitude larger than the thermal energy $k_BT$ at room temperature $T=293K$, while the energy dissipated by the active force along a distance of one diameter of particle $W_a \approx 3\times10^{2}k_BT$ (see Supplementary Information, SI), and the gravitational energy for a change of height of one particle diameter $E_g=7k_BT$ --- the gravitational length is $l_g=0.15\sigma$. The strong dipolar interactions mean that for sufficient field strength, particles form strings reminiscent of those observed in passive dipolar systems \cite{Fraden1989,Yethiraj2003}, although these are active. This string formation is the microscopic origin for electrorheological fluids, \textit{i.e.} a colloidal fluid that gains shear thinning behaviour when placed under an electric field \cite{Klass1967, Gast1989} , and allows to build colloidal polymer chains with tunable stiffness \cite{Vutukuri2012}.

\textit{Overall Phase Behaviour --- }
Consistent with expectations, at low volume fraction, we find a 3d ``active gas'', where the activity is in the $xy$ plane (Fig.~\ref{figPhaseDiagram}b and Supplementary Movie 1 \cite{myWebPage}). When the field strength exceeds the thermal energy, strings form and if the number density of particles is high enough, chains span the entire height of the sample (Fig.~\ref{figPhaseDiagram}c). The symmetry is thus broken in the sense that the $xy$ plane presents a variety of different phases (depending on field strength and colloid volume fraction), whereas the profile in the vertical direction is rather uniform. These strings can interact with each other depending on their number density. At very low density, the separation is  large enough that strings barely interact. These strings then travel in the $xy$ plane, where their average velocity is $\sim 1\mathrm{\mu m/s}$ at $0.3\mathrm{V/\mu m}$, presumably due to the un-aligned orientation of the particles in the strings so that there is some cancellation of the active forces, and which is 2 orders of magnitude much smaller than for individual particles as can be estimated by linear extrapolation from data in \cite{Yan2016} (Supplementary Movie 2 \cite{myWebPage}).

When the volume fraction is increased above $\phi \approx 10^{-2}$, the interaction between the strings leads to self-organisation into sheets (Fig.~\ref{figPhaseDiagram}d). In our system these are active and exhibit a remarkable dynamical behaviour with strong fluctuations in the $xy$ plane (see Supplementary Movie 3 \cite{myWebPage}). When the volume fraction reaches around $\phi\approx0.15$, we find that the active sheets percolate. The resulting structure is reminiscent of a labyrinth (Fig.~\ref{figPhaseDiagram}e), but, as shown in Supplementary Movie 4 \cite{myWebPage}, the system is dynamic and reorganises over time, with opening and closing pathways through the maze (see Supplementary Movie 7 \cite{myWebPage}). At higher volume fraction still ($\phi \approx 0.3$), we find an assembly into a body-centered tetragonal (BCT) crystal (see Supplementary Movie 5 \cite{myWebPage}), and above $\phi\approx0.45$ to a crystal with local hexagonal symmetry which is a mixture of FCC and HCP (see Supplementary Movie 6 \cite{myWebPage}). This crystallisation is reminiscent of passive systems with similar dipolar interactions \cite{Yethiraj2003, Yethiraj2004}. In the next sections, we will focus on the phase behaviour and dynamics of the dense phases ($\phi>0.1$), where interactions between the strings become important and give rise to an unusual dynamical behaviour.

\begin{figure*}
\centering
\includegraphics[width=\textwidth]{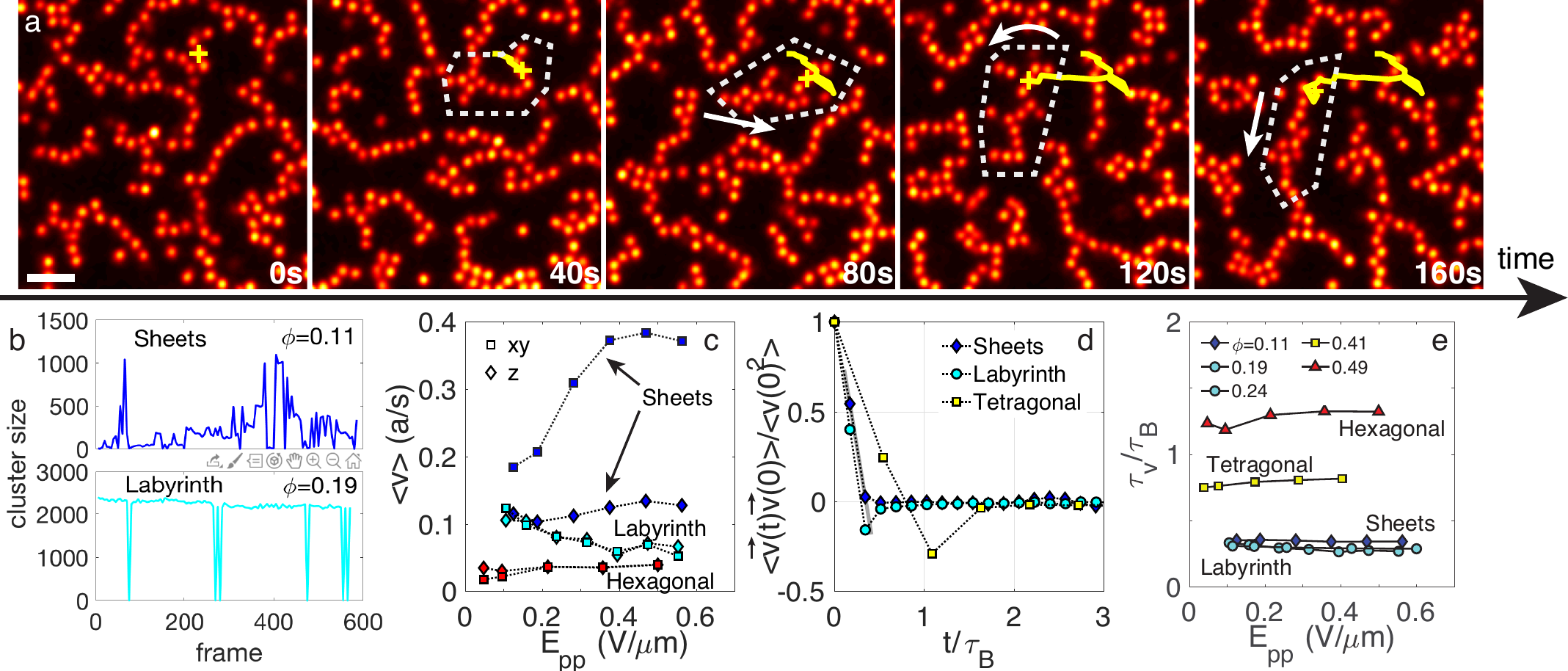}
\caption{\textbf{Dynamic behaviour of Active sheets and Labyrinth. --- }
(a) Successive images of the active sheet phase showing the $xy$ plane. The trajectory of one particle is indicated in yellow. One can notice a cluster of particles detaching from a sheet, travelling in space and collapsing with an another sheet -- highlighted with white dotted lines and arrows. Scale bar corresponds to 5 microns.
(b) Time evolution of cluster size experienced by one particle. Fluctuations of cluster size are large in the sheet phase (top), meaning that aggregation and fragmentation both continuously occur, but are almost absent in the labyrinth (bottom). 
(c) Mean velocity of particles with respect to the electric field (in diameters per second). For the non-percolated active sheet phase, the velocities in the $xy$ plane and $z$ direction are very different, whereas they are almost the same for the percolated labyrinth and body centered tetragonal phases, showing that the motility is transferred from the $xy$ active plane to the $z$ diffusive direction. Note that the velocity for the labyrinth phase decreases with the electric field. 
(d) Velocity autocorrelation for different phases. For the labyrinth and tetragonal crystal phases, a dip at finite time with negative correlation appears, whereas it is absent for the active sheet phase. The $x$-axis is normalized by the Brownian time $\tau_B$ at dilute limit. 
The grey line corresponds to the extrapolation to obtain the decorrelation time of the velocity in the sheet phase. 
(e) Velocity decorrelation time $\tau_v$ for sheets, labyrinth, tetragonal and hexagonal phases with respect to electric field. $\tau_v$ is defined as the first passage time to zero of the autocorrelation functions, except for the sheet phase where we find $\tau_v$ by extrapolation of the first data points.
}
\label{figDynamicsVelocity} 
\end{figure*}

\textit{Dynamic behaviour of Active sheets and Labyrinthine Phases --- } The active sheet state is characterised by a large fluctuating dynamical behaviour with breaking and reforming of links between sheets. Such events are indicated in Fig. \ref{figDynamicsVelocity}(a), where an assembly of particles detaches from a sheet, travels, and sticks to an another sheet. That is to say, particles detach from one sheet and reattach to another so that the size of the sheet they belong to changes all the time. Indeed, in Fig.~\ref{figDynamicsVelocity}(b), we display the time evolution of the size of the \textit{cluster} that one arbitrary particle belongs to. We emphasise that this is an extensive quantity, characterising the local connectivity of the particles. Here a sheet is defined as a assembly of particles connected together with a maximum distance of $1.2$ diameters, which is the first minimum of the pair correlation function. We can see that in the sheet phase, the size changes continuously, jumping from a very low value to a value close to the total number of particles. This results from a population of unattached particles which periodically attach to the cluster which comprises the vast majority of particles in the system. On the contrary, in the labyrinthine phase, the cluster size is close to the total number of particles in the volume sampled except occasionally dropping to one, indicating a rare jump event occurring.

Fig.~\ref{figDynamicsVelocity}(c) shows the mean velocity of the particles -- in both the $xy$ plane and $z$ direction -- in the sheet, labyrinthine and tetragonal crystal phases. In the sheet phase, we find a strong anisotropy in the velocity with the lateral velocity in the $xy$ plane is much higher than in the $z$--direction. This reflects the fact that the sheets move laterally (recall that the particles are motile in the $xy$ plane), but not vertically (see Supplementary movie 3 \cite{myWebPage}). Whereas for the labyrinth and tetragonal phases, percolation leads to a drastic drop in the $xy$ velocity with respect to the sheets [Fig. \ref{figDynamicsVelocity}(c)]. We see that in the labyrinth phase, the velocity decreases with the electric field.

We now consider the velocity autocorrelation
\begin{equation}
c(t) = \frac{\left< \vec{v}_i(t_0+t)\vec{v}_i(t_0)\right>_{i,t_0}}{\left<||\vec{v}_i(t_0)||^2\right>_{i,t_0}}
\end{equation}
\noindent
where $\vec{v}_i$ is the instantaneous velocity of particle $i$ and $\left<\cdot \right>_{i,t_0}$ denotes an average over all particles $i$ and initial time $t_0$. Interestingly, the labyrinth and crystal phases present a dip below zero which is absent for the sheet phase. This suggests that there is an emergence of elasticity in the system, which is consistent with the expectation that percolation leads to elastic solidity, even in this active system. In Fig. \ref{figDynamicsVelocity}(e), we consider the timescale over which the velocity correlations are lost. Interestingly, this increases with volume fraction, and does not depend on the field intensity for all phases.

\begin{figure*}
\centering
\includegraphics[width=0.8\textwidth]{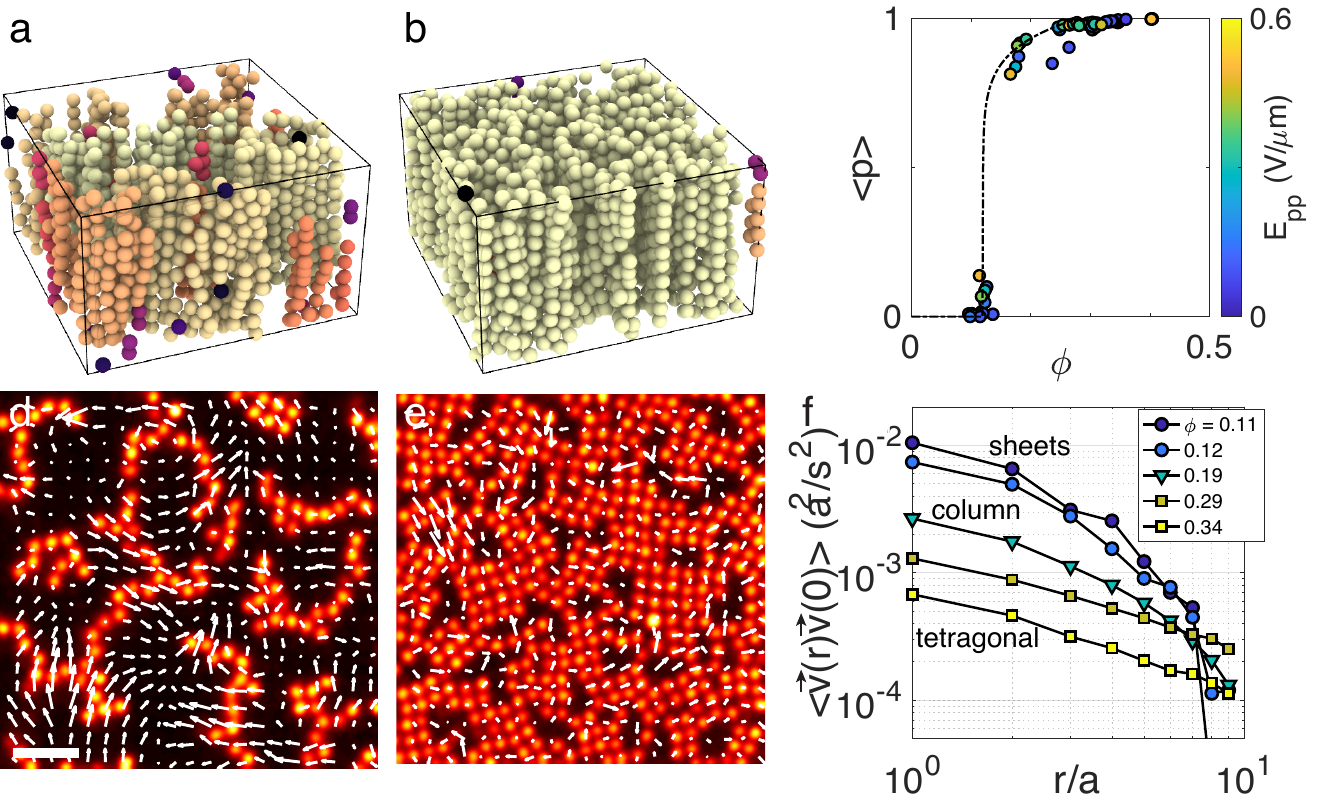}
\caption{\textbf{Emergence of the Dynamic labyrinth --- } (a,b) 3d rendered image of the suspension, where each particle is coloured depending on the aggregate of connected particles it belongs to, at (a) $\phi=0.11$ and (b) $0.19$. 
(c) Percolation order parameter (defined in the text) with respect to volume fraction. Colour of the symbols correspond to the electric field applied to the sample. 
(d,e) 2d xy slices of confocal images, with local velocity field obtained from an image regression algorithm (see text) in the sheet (d) and labyrinth (e) phases under $E_{pp} = 0.26V/\mu m$. Scale bar corresponds to 5 microns.
(f) Velocity spatial correlation functions for sheets ($\circ$), labyrinths ($\triangledown$) and tetragonal ($\square$) phases under $E_{pp} = 0.15\mathrm{V/\mu m}$. The function correlates the velocity - in diameter per second a/s - of two particles in the same $xy$ plane \textit{i.e.} with distance in the $z$ direction less than half diameter.}
\label{figPercolationClusterDynamics}
\end{figure*}

\textit{Emergence of the Labyrinthine Phase --- }
We now consider the percolation transition between the active sheets and dynamic labyrinth. Figure~\ref{figPercolationClusterDynamics}(a,b) shows two rendered 3d images of the system
at a volume fraction of $\phi=0.11$ and $\phi=0.19$ respectively, where particles are coloured depending on the percolating cluster they belong to. At $\phi=0.11$, the sheets span in the direction parallel to the electric field, but are separated in the $xy$ plane whereas at $\phi=0.19$, the system percolates (at least on the lengthscale of our images). The change in the structure can be captured using the standard order parameter for percolation \textit{i.e.} the probability $p$ of a given particle to belong to the ``infinite'' cluster, where we define the infinite cluster as the largest cluster that spans the sample and connects the six boundaries of the 3d image. This quantity is averaged in time, and is plotted in Fig.~\ref{figPercolationClusterDynamics}(c). The time average order parameter $\left<p\right>$ is zero for densities less than $0.15$ and non zero for larger densities. Moreover, there is a clear discontinuity to the slope of the order parameter at $0.15$, which is zero below $0.15$ and close to infinite above, which is a signature of a percolation transition. Notice that the region sampled in our images is not large enough to determine precisely the percolation volume fraction, and we thus have here an estimate of where percolation occurs  in our system.

The percolating structure is somewhat reminiscent of a 2d network in the $xy$ plane which is replicated in the $z$ direction, but here we find that the system is highly dynamic and indeed pathways through the labyrinth open and close. Figure ~\ref{figPercolationClusterDynamics}(d,e) shows typical particle velocity fields in the $xy$ plane for the sheets (d) and labyrinth (e). At volume fractions less than percolation [Fig.~\ref{figPercolationClusterDynamics}(d)], the velocity is correlated in space with several vortices whose length scale is around ten particle diameters. At $\phi=0.19$ [Fig.~\ref{figPercolationClusterDynamics}(e)], the suspension has percolated and the velocity field becomes more random and isotropic, the average velocity becomes much smaller of the order of $0.1 \mathrm{a/s}$ [Fig.~\ref{figDynamicsVelocity}(c)], with some regions where vortices persist, and an emergence of an apparent dynamical heterogeneity. To quantify the change in cooperative motion, we computed the spatial velocity correlation $\left<\vec{v}(r)\vec{v}(0)\right>$ between two particles in the same $xy$ plane [Fig.~\ref{figPercolationClusterDynamics}(f)] where $r=x^2+y^2$. Interestingly, the spatial correlation tends to extend to larger distances when the phase changes from the sheets at low density to the crystal at higher density. Moreover, while the velocity correlations for the sheets and labyrinthine phase seem to decrease exponentially, the decrease for the crystal seems to follow a power law, but the range of the correlations is not long enough to settle this definitively.

\begin{figure*}
\centering
\includegraphics[width=\textwidth]{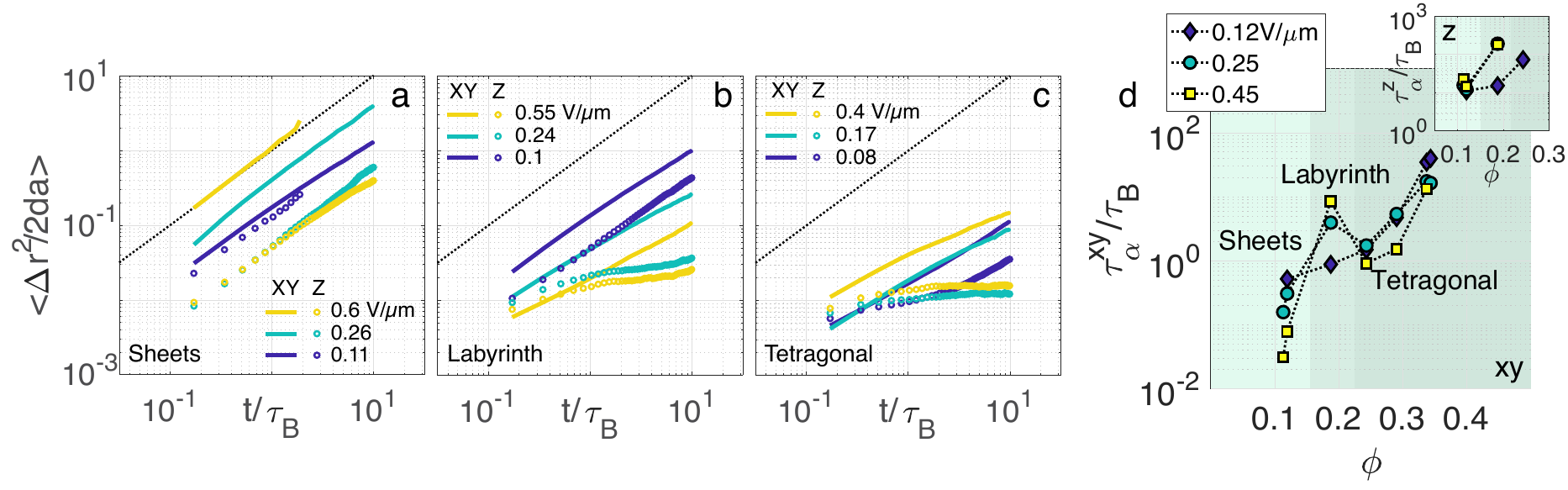}
\caption{\textbf{Diffusive dynamics and particle transport ---} (a-c) Mean square displacements (MSD) in xy (orthogonal) and z (parallel) direction for sheet (a), labyrinth (b) and tetragonal (c) phases. The MSD is normalized by $2da$ where $d$ is the dimension and $a$ is the diameter of particles, and the time is normalized by the brownian time $\tau_B$ in the dilute limit. The dotted lines correspond to the MSD of an isolated particle in the dilute limit. 
(d) $\alpha$--relaxation time -- in units of the Brownian time $\tau_B$--- with respect to volume fraction at low, middle and large electric field. The main panel corresponds to the relaxation time measured from the MSD in xy directions, and the inset corresponds to displacements in z direction.}
\label{figMSDalphaTime}
\end{figure*}

\textit{Mean square displacement --- }
This complex dynamical behaviour leads to different transport regimes depending on the phase of the system. Figures~\ref{figMSDalphaTime}(a-c) show the mean square displacement for the active sheets, labyrinth and tetragonal crystal respectively. First, the mean square displacement is larger in the $xy$ plane than in $z$--direction, which is consistent with the fact that activity is directed in the $xy$ plane. Second, the active sheet phase exhibits a diffusive behaviour both in $xy$ and $z$ directions, whereas it becomes subdiffusive for the labyrinthine and the crystalline phases. Indeed, when the system percolates, most of the particles are bound. 
Motion is thus correlated in space, which is consistent with the increase of the spatial extent of the velocity correlation Fig.~\ref{figPercolationClusterDynamics}. It is interesting to see that even in the tetragonal crystal, particles still diffuse at rather short timescales. Third, and importantly, the particle transport increases or decreases with the field depending on the phase of the system: the sheets and the crystal shows an increase of the transport with the electric field, while the labyrinths shows the opposite. This can be explicitly quantified by computing the self-intermediate scattering function $F(k,t)$ and measuring the $\alpha$--relaxation time $\tau_\alpha$, which for a suitable choice of wave-vector $k$ quantifies the time needed for one particle to move a distance of one diameter (see SI).

Figure~\ref{figMSDalphaTime}(d) shows the relaxation time for particle transport in $xy$ and $z$ respectively. First and generally speaking, the relaxation times increase with the volume fraction. Then, in the $xy$ plane [Fig.\ref{figMSDalphaTime}(d)], while the sheet phase exhibits a substantial decrease of the relaxation time \textit{i.e.} an acceleration of the dynamics with the field, the labyrinth sees a \emph{decrease} of the dynamics of the same order of magnitude. Then, in the tetragonal phase we see again an acceleration of the dynamics with the field. For the dynamics in the $z$ direction [Fig.\ref{figMSDalphaTime}(d) inset], we can only measure the relaxation time at relatively low density where the relaxation takes place in a range of time accessible to the experiments. In these cases, the relaxation time always increases with the electric field, except the sheet phase which is rather insensitive to the field strength.

\begin{figure}
\centering
\includegraphics[width=0.5\textwidth]{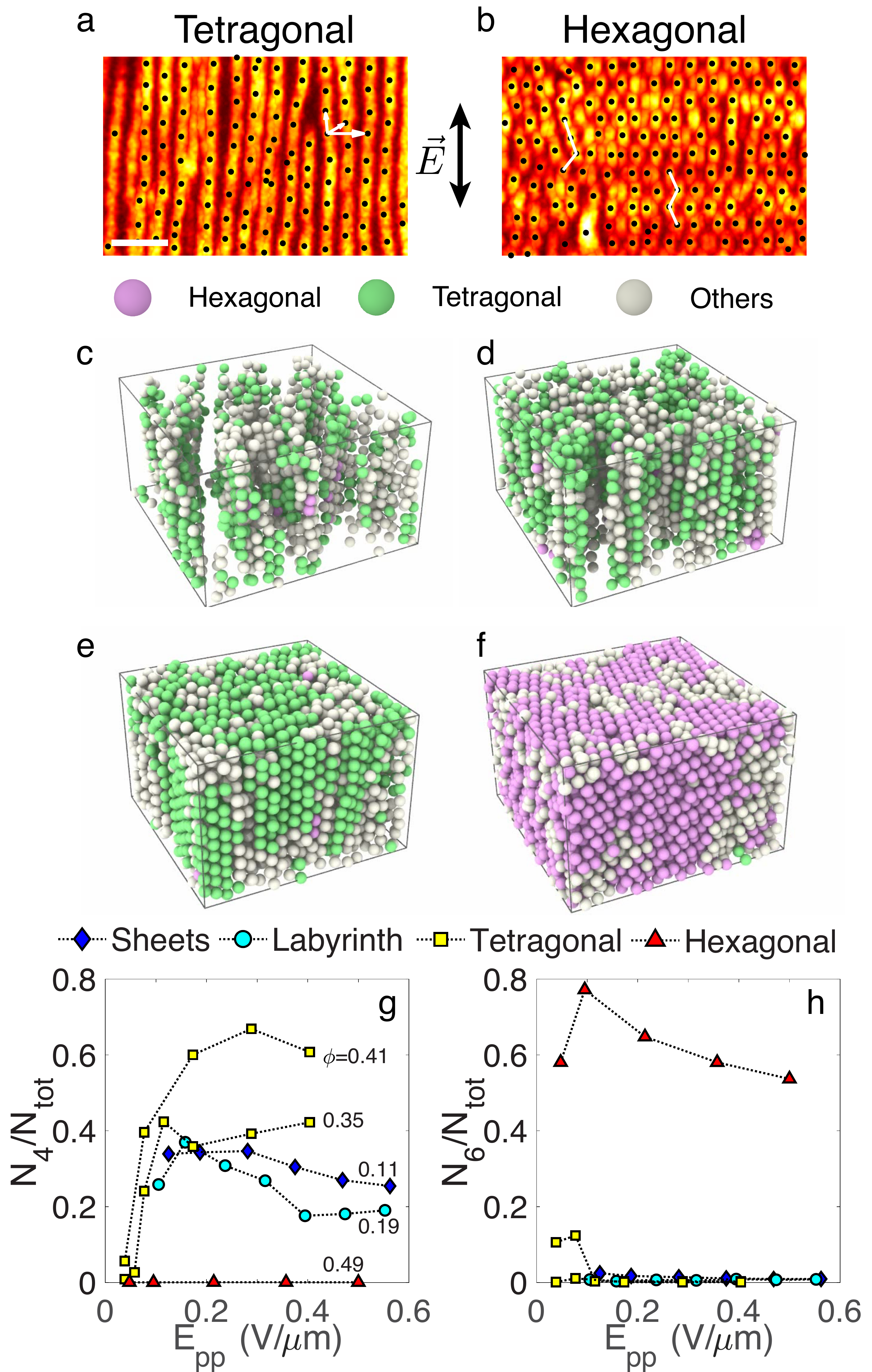}
\caption{\textbf{Crystal structure --- } 
(a,b) 2d $xz$ slices of the 3d confocal images of tetragonal and hexagonal phases. Tracking results are superimposed and the different crystal structures are highlighted in white. One can notice both FCC and HCP structures are present in the hexagonal phase. Scale bar corresponds to 5 microns.
(c-f) 3d rendered images obtained from tracking results of respectively sheets, labyrinth, tetragonal and hexagonal phases. Colours correspond to the local crystalline order obtained from $\bar{q}_4$ and $\bar{q}_6$ as explained in the text. 
(g,h) Time averaged crystal populations $N_{4,6}$ with respect to electric field $E_{pp}$ in different phases. The crystal populations are normalized by the total number of particles $N_{tot}$.}
\label{figCrystalOrder}
\end{figure}

\textit{Crystallisation: cubic and hexagonal local structure --- }Finally, let us consider the crystalline order appearing at high density. Figure~\ref{figCrystalOrder}(a,b) shows two-dimensional slices in the $xz$ plane of confocal images in the cubic and hexagonal phase respectively. Two neighbouring strings in the tetragonal phase are shifted in the $z$--direction at about half a particle diameter. In the hexagonal phase, the entropic interactions favour a hexagonal structure and the strings disappear. We see in Fig.~\ref{figCrystalOrder}b that both HCP and FCC structures are found in the hexagonal crystal. To capture the increase of crystalline order, we defined for each particle $i$ two quantities $\bar{q}^i_{4},\bar{q}^i_{6}$ based on quasi-2d bond orientational order parameter and first neighbour correlation \cite{Yethiraj2004,Bialke2012} (see SI). These quantities are constructed such that they range from $-1$ to $1$. Particles in a locally disordered structure have $\bar{q}_4=0,\bar{q}_6=0$, whereas particles in a tetragonal or hexagonal structure have $\bar{q}_{4,6}=1$ \cite{Bialke2012}. We can thus attribute each particle a local structure by thresholding: particles having $\bar{q}_{4}>0.5$ and $\bar{q}_{4} > \bar{q}_{6}$ are taken to be in a tetragonal structure, whereas $\bar{q}_{6}>0.5$ and $\bar{q}_{6} > \bar{q}_{4}$ are in a hexagonal structure.

Figure~\ref{figCrystalOrder}(c-f) shows rendered 3d images from sheet to hexagonal phases where particles are coloured according to the local crystal structure, and show the appearance of crystal nuclei in the tetragonal and hexagonal phases. We see that the sheet and labyrinth phases have a non-negligible number of particles in a tetragonal structure of about 35\%. Moreover, some particles also have a local hexagonal structure, and the hexagonal phase also has a few particles in a tetragonal structure. To quantify the dependence with the density and the electric field, we define the crystal population $N_4,N_6$ as the number of particles respectively in a local tetragonal and hexagonal structure. These quantities are plotted in Fig.~\ref{figCrystalOrder}(g-h), and show that tetragonal crystal population increases with the electric field from zero to reach a stationary value at $E_{pp}\sim0.15\mathrm{V/\mu m}$, while hexagonal crystal population is non-zero and is almost independant whatever the electric field. Interestingly, strings in phases at lower volume fraction (columns, sheets and labyrinths) also appears above $E_{pp}\sim0.15\mathrm{V\mu m}$. All these suggest that, similarly to the passive case \cite{Yethiraj2003}, the hexagonal structure is governed by packing in contrast with the tetragonal structure that is determined by dipolar energy.

\textit{Discussion --- }The 3d active system that we have realised exhibits a rich phase behaviour with number of unexpected features based on predictions from computer simulations and theory of active colloidal systems, in 3d \cite{Cates2013, Stenhammar2014, Wysocki2014,Winkler2015} and the passive analogue of this active dipolar system \cite{Yethiraj2003,Yethiraj2004}. Firstly, the self-assembly of a sheet phase is a surprise and is unexpected given the direct interactions. It is possible that our assumption that the dipole in each hemisphere is of similar magnitude \cite{Yan2016} is insufficient. In addition to operating in 3d, there are some differences to the earlier work \cite{Yan2016}, for example the thickness of the metal layer (ours is ten times thinner). However, it is not clear why this should promote the symmetry breaking that leads to sheet formation. We speculate that this is caused by some electrohydrodynamic interaction which may be related to similar structures seen in strongly confined 2d dipolar systems \cite{Gong2002}. The P\'{e}clet number that our system reaches is comparable with that required for motility induced phase separation in the case of active Brownian particles in 3d, as is the volume fraction \cite{Wysocki2014,Stenhammar2014}. Of course the interactions are different here, due to the dipolar contribution, but we also note that our system size here is 30 particle diameters between the electrodes, and finite size effects have been found to be very significant in computer simulations of active systems \cite{Stenhammar2014}.

Secondly, the labyrinth phase has rather striking characteristics. Formed from the percolation of the sheets, its structure is reminiscent of a 2d network replicated in the $z$ direction, the dynamics is rather closer to that of a liquid, albeit with a sub-diffusive behaviour. This is due to the fact that the structure is changing all the time: sheets and labyrinths are fragmenting and aggregating constantly, and particles jump from one another and constantly change neighbours, even at relatively high volume fraction. This is somewhat reminiscent of cytoskeletal active matter \cite{Sanchez2012, Guillamat2017}, where microtubules are attached together and slide past each other thanks to molecular motors. But in contrast with the microtubules, the orientation of the motion of constituents are \textit{a priori} not imposed by the structure of our active network, and microtubules possesses an inherent stiffness in contrast with our sheets and strings.

Finally, we found an unexpected non-monotonic dependance of the transport dynamics with respect to the volume fraction. Counterintuitively, the dynamics in the labyrinthine phase is slower than in the tetragonal phase, despite its lower volume fraction and having an open structure which is only ordered in the $z$-direction. Moreover, the electric field effects the dynamics in opposite ways, depending on the phase of the system. The dynamics decrease with the electric field in the labyrinthine phase while accelerating for the other phases, which enhances the non-monotonicity raised above. This is reminiscent of recent work with active glasses where the state point can affect dramatically the dynamical behaviour \cite{Berthier2017, Klongvessa2019}. Keeping in mind that the electric field increases both the activity \textit{and} the dipolar binding, our result suggests that one effect dominates to the other depending on the phase. This is all the more surprising given that the dipolar interaction is the predominant energy scale compared to the activity whatever the field strength.

\textit{Conclusions. --- }
In conclusion, we explored experimentally the phase behaviour of a 3d suspension of active colloids. The observation has been carried out using a confocal microscope, providing access to microscopic details in space and in time. The competition between dipolar interaction and activity gives rise to a complex and rich behaviour, ranging from an isolated aggregate phase and an active dense disordered phase. In particular, we uncovered an unexpected symmetry breaking to form an active sheet phase with large dynamical fluctuations. This could allow one to consider these sheets as a model experimental system to study biological membrane fluctuations \cite{Turlier2018}. These sheets can then self-assemble to form a rather unusual labyrinthine phase, where particles form a structure reminiscent of a 2d network extended in the third dimension, but where activity leads the particles to jump from one branch to an another, and diffuse in the percolating network. It may be interesting to investigate how this structure responds to and whether or not it can sustain a mechanical stress. In closing we note that this work also opens new possibilities to study active matter in 3d at high density, where we can enquire how activity can affect phenomena such as percolation crystallisation and glass transition \cite{VanDerMeer2016, Berthier2017, Klongvessa2019} and phenomena such as ``bubbly'' phase separation \cite{Tjhung2018}. Finally, we note that we have only focussed on one frequency of the applied field. Given the behave of this class of Janus particles in 2d \cite{Yan2016}, one can expect more exotic behaviour by going to higher frequency.

\begin{acknowledgments}
The authors would like to thank Francesco Turci, James Hallett, Masaki Sano and Tannie Liverpool for helpful discussions and Anton Souslov for a critical reading of the manuscript.
N.S. would like to thank Andrew Murray, Germinal Magro and Jean-Charles Eloi for technical support.
The authors would like to acknowledge the European Research Council (FP7 / ERC Grant agreement n$^\circ$ 617266 "NANOPRS'') and the Engineering and Physical Sciences Research Council (EPH022333/1) for financial support.
\end{acknowledgments}

\bibliographystyle{apsrev4-1}
\bibliography{2019spider.bib}

\end{document}